\documentclass[%
 reprint,
 amsmath,amssymb,
 aps,
]{revtex4-2}

\usepackage{graphicx}
\usepackage{dcolumn}
\usepackage{bm}
\usepackage[many]{tcolorbox}
\usepackage{lipsum}
\usepackage{amsmath}
\usepackage{shuffle}
\usepackage{tikz}
\usepackage{environ}
\usepackage{tabularx}
\usetikzlibrary{shapes.geometric, arrows,decorations}

\makeatletter
\newsavebox{\measure@tikzpicture}
\NewEnviron{scaletikzpicturetowidth}[1]{%
  \def\tikz@width{#1}%
  \begin{lrbox}{\measure@tikzpicture}%
  \BODY
  \end{lrbox}%
  \pgfmathparse{#1/\wd\measure@tikzpicture}%
  \BODY
}
\makeatother
\begin{document}

\title{Universal expansions of scattering amplitudes\\
for gravitons, gluons and Goldstone particles}

\author{Jin Dong$^{1,4}$\footnote{dongjin@itp.ac.cn}, Song He$^{1,2,3,4,5}$\footnote{songhe@itp.ac.cn} and Linghui Hou$^{1,2,4}$\footnote{houlinghui@itp.ac.cn}}
\affiliation{
$^{1}$CAS Key Laboratory of Theoretical Physics, Institute of Theoretical Physics, Chinese Academy of Sciences, Beijing 100190, China \\
$^{2}$School of Fundamental Physics and Mathematical Sciences, Hangzhou Institute for Advanced Study, UCAS, Hangzhou 310024, China \\
$^{3}$International Centre for Theoretical Physics Asia-Pacific, Beijing/Hangzhou, China\\
$^{4}$School of Physical Sciences, University of Chinese Academy of Sciences, No.19A Yuquan Road, Beijing 100049, China\\
$^{5}$Peng Huanwu Center for Fundamental Theory, Hefei, Anhui 230026, P. R. China
}\date{\today}

\begin{abstract}
Tree-level scattering amplitudes for gravitons, gluons and Goldstone particles in any dimensions are strongly constrained by basic principles, and they are intimately related to each other via various relations. We study two types of ``universal expansions" with respect to gauge bosons and Goldstone bosons: the former express tree amplitudes in Einstein gravity (Yang-Mills) as linear combinations of single-trace Einstein-Yang-Mills (Yang-Mills-$\phi^3$) amplitudes with coefficients given by Lorentz products of polarizations and momenta; the latter express tree amplitudes in non-linear sigma model, (Dirac-)Born-Infeld and a special Galileon theory, as linear combinations of single-trace mixed amplitudes with particles of lower "degree of Adler's zero" and coefficients given by products of Mandelstam variables. We trace the origin of gauge-theory expansions to the powerful uniqueness theorem based on gauge invariance, and expansions in effective field theories can be derived from gauge-theory ones via a special dimension reduction.
\end{abstract}

\maketitle


\section{Introduction}
Recent years have witnessed enormous progress in unravelling unexpected simplifications and new, hidden mathematical structures for scattering amplitudes in Quantum Field Theory (c.f.~\cite{Elvang:2013cua}). Such remarkable structures have been found not only for amplitudes in special theories such as ${\cal N}=4$ super-Yang-Mills to all loop orders~\cite{Arkani-Hamed:2010zjl,Arkani-Hamed:2012zlh,Arkani-Hamed:2013jha}, but also for tree amplitudes in a wide range of theories such as gauge theories, gravity and effective field theories (EFTs)
. Moreover, deep, universal relations have been discovered for such amplitudes of gluons, gravitons, and Goldstone particles ({\it c.f.}~\cite{Bern:2008qj, CHY4, Cheung:2017ems, Cheung:2017pion}), and the origin for some of these relations still remain to be understood. 

Tree amplitudes in general relativity (GR) and Yang-Mills theory (YM), which encode leading two-derivative interactions of gravitons and gluons, turn out to be uniquely determined by gauge invariance, provided that one starts with an ansatz of cubic tree graphs and correct power counting~\cite{Arkani-Hamed:2016rak,Rodina:2016jyz}.
This is rather remarkable as it implies that unitarity and locality, reflected in factorization of amplitudes, can be derived from gauge invariance and singularity structures for these amplitudes. On the other hand, Goldstone particles for spontaneous symmetry breaking have intriguing infrared behavior encoded in soft limits~\cite{Kampf:2013vha,Cheung:2014dqa}: certain amplitudes of these EFTs have enhanced Adler zero which are totally invisible in Feynman diagrams, similar to gauge invariance for gauge theories and gravity. These include pions in non-linear sigma model (NLSM), scalars in Dirac-Born-Infeld (DBI) and even a special (most symmetric) Galileon theory (sGal), with increasingly vanishing soft behavior~\cite{Cheung:2014dqa,Cheung:2015ota,Cheung:2016drk} known as Adler zero~\cite{Adler:1964um}. Similarly, with quartic-graph ansatz and correct power counting, these amplitudes are uniquely determined by enhanced Adler zero under soft limits~\cite{Rodina:2016jyz}. Born-Infeld (BI) amplitudes enjoy both gauge invariance and Adler zero, though it is slightly more non-trivial to fix them using such conditions~\cite{Cheung:2018oki}.

All these amplitudes are closely related to each other via a web of relations. Perhaps the most famous ones are double-copy relations (see the review~\cite{Bern:2019prr} and references therein), originally discovered by KLT in string theory~\cite{Kawai:1985xq} and by BCJ via color-kinematics duality in QFT~\cite{Bern:2008qj,Bern:2010ue}, which can be summarized in the slogan ``GR$=$YM $\otimes$ YM"~\footnote{To be more precise, double copy of YM gives gravity coupled to B-field and dilaton.}. We use $L \otimes R$ to denote the theory whose amplitudes are obtained by field-theory KLT~\cite{Bern:1998sv, Bjerrum-Bohr:2010pnr} or equivalently tree-level BCJ double-copy, of amplitudes in theories $L$ and $R$ with color/flavor structures; this $\otimes$ operation becomes particularly natural and universal in Cachazo-He-Yuan (CHY) formulas~\cite{CHY1,CHY2,CHY3,CHY5,CHY4}, and by construction bi-adjoint $\phi^3$ serves as the identity~\cite{CHY3}: $I=I \otimes \phi^3=\phi^3 \otimes I$ for any theory $I$. From this perspective, it has been extended to a large class of theories including EFTs, {\it e.g.} (D)BI$=$ YM(s) $\otimes$ NLSM, and sGal $=$ NLSM $\otimes$ NLSM~\cite{CHY4}. Two more operations connect these amplitudes in theory space, which can be understood from CHY formulas~\cite{CHY5, CHY4} or more directly as ``unifying relations"~\cite{Cheung:2017ems}. The $\oplus$ operation produces {\it mixed amplitudes} of two types of particles, $I \oplus II$, with particular interactions between them, and additional color/flavor structure for $II$ (compared to $I$), {\it e.g.} GR $\oplus$ YM $=$ Einstein-Yang-Mills (EYM) and an interesting ``extended DBI" theory from DBI $\oplus$ NLSM~\cite{CHY4} (see also \cite{Kampf:2021tbk}). 
A special dimension reduction (DR) ``reduce" gauge bosons to Goldstone bosons~\cite{Cheung:2017pion}, {\it e.g.} YM to NLSM (gluons to pions), GR to BI (gravitons to photons) {\it etc.} 

In this note, we study a different type of relations among amplitudes which further demonstrate how these amplitudes are strongly constrained and closely related to each other. We call them ``universal expansions" since they apply to amplitudes of gravitons, gluons and Goldstone particles universally, and they encode $\oplus$, $\otimes$ as well as DR in a natural way. The amplitude with $n$ gravitons can be expanded as a linear combination of EYM (GR $\oplus$ YM) mixed amplitudes with $r{+}2$ gluons in a single trace and $n{-}r{-}2$ gravitons (for $r=0, 1, \cdots, n{-}2$), and each coefficient is a Lorentz product of $r{+}2$ polarization vectors and $r$ momenta; the same holds if we replace gravitons (gluons) with gluons(bi-adjoint $\phi^3$ scalars) in YM, or photons in BI (pions in NLSM). Note that these mixed amplitudes, which turn out to be building blocks of the original amplitude, are simpler since the new particles have spin lowered by one compared to the original ones. Remarkably, we find that with ansatz of~\cite{Rodina:2016jyz}, gauge invariance in $n{-}1$ particles uniquely determines this expansion, thus indirectly fixes all these mixed amplitudes! More precisely, gauge invariance in $n{-}2$ gravitons/gluons already fixes the form of the expansion, with each term obtained by certain differential operators acting on the full ansatz~\cite{Rodina:2016jyz}; by imposing gauge invariance on any of the remaining two particles, we uniquely fix the expansion and consequently these mixed amplitudes. Since EFT amplitudes can be obtained via special DR from gauge-theory ones, they have similar expansions {\it e.g.} BI (NLSM) amplitude as a linear combination of mixed amplitudes of BI $\oplus$ YM (NLSM $\oplus$ $\phi^3$), with coefficients given by products of Mandelstam variables. The new particles in these mixed amplitudes, gluons ($\phi^3$ scalars), have lower degree of Adler zero (by two) compared to that of the original ones, photons (pions).

Let us summarize our main results as follows. For gauge theories and gravity, their amplitudes satisfy gauge invariance: $A_n$ is invariant under $e^\mu_i \to e^\mu_i +\alpha~p^\mu_i$ for $i=1,\cdots n$. It turns out that gauge invariance implies an expansion for the $n$-point tree amplitude of theory $I$~\footnote{For YM amplitude, we always refer to partial amplitude with ordering $(1,2,\cdots, n)$; similarly for NLSM and one of the two orderings for bi-adjoint $\phi^3$ case.}:
\begin{equation} \label{expand one}
	{A}_n^{\uppercase\expandafter{\romannumeral1}}= \sum_{\alpha} (-1)^r W(1,\alpha,n) A^{\uppercase\expandafter{\romannumeral1} \oplus \uppercase\expandafter{\romannumeral2}}( \{\bar{\alpha}\}|1,\alpha,n ),
\end{equation}
where we single out two special legs, {\it e.g.} $1, n$; and the sum is over all {\it ordered subsets} of $\{2,\cdots,n-1\}$ denoted by $\alpha$ (with $|\alpha|=r$ for $r=0, 1, \cdots, n{-}2$), and $\{\bar{\alpha}\}$ denotes the complementary (unordered) set with $n{-}2{-}r$ labels~\footnote{The number of terms equal the total number of all permutations of $n{-}2$ objects, $\sum_{i=0}^{n{-}2} \frac{(n{-}2)!}{i!}$.}. For each term, we have a Lorentz-contraction prefactor
\begin{equation}
    W(1,\alpha,n)\equiv  e_1 \cdot f_{\alpha_1} \cdot f_{\alpha_2} \ldots  f_{\alpha_r} \cdot e_n,\nonumber
\end{equation}
with linearized field-strength $f_i^{\mu\nu}\equiv p_i^\mu e_i^\nu-e_i^\mu p_i^\nu$, and a mixed amplitude of $n{-}2{-}r$ particles in theory $I$ in $\{\bar{\alpha}\}$ and $r{+}2$ particles in theory $II$ with spin lowered by one; they are ordered as $(1,\alpha,n):=(1, \alpha_1, \cdots, \alpha_r, n)$ since they carry additional color/flavor structure. These theories are summarized in the Table as follows. 
\begin{table}[htbp]\label{one}
\begin{tabular}{|c|c|c|c|}  
\hline
&\multicolumn{3}{c|}{1. Gauge Theories} \\  
\hline  
I & GR ($s=2$)  & YM ($s=1$) & BI ($s=1$)\\  
\hline  
II & YM ($s=1$) & $\phi^3$ ($s=0$) & NLSM ($s=0$)\\  
\hline  
\end{tabular} 
\end{table} 

There are similar universal expansions for EFT amplitudes with (enhanced) Adler zero; the defining properties of these EFTs is that their amplitudes vanish (to certain a degree) under soft limit: for $p_i^\mu = \tau \hat{p}_i^\mu$ with $\tau \to 0$, we have $\lim_{p_i\sim {\cal O}(\tau) \to 0}A_n= \mathcal{O}(\tau^h)$ for any $i$,
where we call $h$ the degree of Adler zero. These amplitudes and their expansions follow from gauge theory and gravity ones via a special DR. One way for expanding amplitudes in theory $I$ with $n$ Goldstone bosons is  
\begin{equation} \label{expand two}
	{A}_n^{I}=\sum_{\alpha; \mathrm{odd}} \hat{W}(1,2,\alpha,n) \; A^{I \oplus II}( \{\bar{\alpha}\}|1,2,\alpha,n ),
\end{equation}
where, after singling out $1,2,n$, the sum is over ordered subsets of $\{3,\cdots, n{-}1\}$ with $r=|\alpha|$ odd, and $$\hat{W}(1,2,\alpha,n) \equiv p_2\cdot \hat{f}_{\alpha_1}\cdots \hat{f}_{\alpha_r}\cdot p_{n}$$ with $\hat{f}^{\mu\nu}_i\equiv p_i^\mu p_i^\nu$, or $\hat{W}=s_{2,\alpha_1} s_{\alpha_1, \alpha_2} \cdots s_{\alpha_r, n}$ (with $s_{i,j}:=p_i \cdot p_j$)~\footnote{Our definition of Mandelstam variables differs by a factor of 2 from the conventional one; such overall constants of amplitudes can be absorbed into the definition of coupling constants which have been stripped off.}; the mixed amplitudes has $(n{-}3{-}r)$ particles of theory $I$ in $\{\bar{\alpha}\}$ and $r{+}3$ particles of theory $II$ with degree of Adler's zero $h$ reduced by two, which carry additional color/flavor structure and are ordered as $(1,2,\alpha,n)$. These theories and their degree of Adler's zero can be found in the following Table~\footnote{In the last column, we refer to pure scalar sector of DBI theory and pure-scalar amplitudes of Yang-Mills-scalar (YMs) theory, which are the dimensional-reduction of BI and YM, respectively.}.
\begin{table}[htbp]\label{two}  
\begin{tabular}{|c|c|c|c|c|}  
\hline                     
& \multicolumn{4}{c|}{2. Effective Field Theories}  \\  
\hline  
\uppercase\expandafter{\romannumeral1}  & sGal ($\tau^3$) & NLSM ($\tau^1$) & BI ($\tau^1$)& DBI ($\tau^2$)\\  
\hline  
\uppercase\expandafter{\romannumeral2}  & NLSM ($\tau^1$)& $\phi^3$ ($\tau^{-1}$) & YM ($\tau^{-1}$)& YMs ($\tau^0$)\\  
\hline  
\end{tabular}  
\end{table}

We remark that these mixed amplitudes encode highly non-trivial interactions between particles in theory $I$ and $II$, and it is not obvious at all why they appear as ``building blocks" of universal expansions of pure amplitudes of theory $I$. For example, the mixed amplitudes of Yang-Mills $\oplus$ bi-adjoint $\phi^3$ come from the Lagrangian in~\cite{CHY4}, which consists of YMs from dimension reduction and the bi-adjoint $\phi^3$ interaction term;
mixed amplitudes of EYM$=$GR $\oplus$ YM encode the well-known minimal coupling of gravitons and gluons. For BI $\oplus$ NLSM and other cases for EFTs, it is only after computing these mixed amplitudes via CHY or DR, can we determine the Lagrangian with rather intricate interaction terms~\cite{Cachazo:2016njl,Mizera:2018jbh}!


Since these mixed amplitudes were originally discovered via CHY, the two sets of expansions can be easily derived using these formulas as well (see~\cite{Fu:2017uzt,Teng:2017tbo, Du:2017kpo, Du:2017gnh} and {\it e.g.} ~\cite{Feng:2019tvb,Zhou:2019mbe}); but this just shifts the question to why these amplitudes have such nice CHY formulas. In this note we do not rely on CHY formulas at all, but instead we find that all mixed amplitudes in gauge theories and gravity are indirectly determined by the uniqueness of the full amplitude in GR/YM, which is naturally written as an expansion. Note that each term in the gauge-theory expansion is gauge invariant with respect to $n{-}2$ particles, which is the best one can achieve for any functions other than the full amplitude~\cite{Rodina:2016jyz,Arkani-Hamed:2016rak}. In a sense, such expansions resemble certain Taylor expansions for the full amplitude: each coefficient is the analog of ${\bf x}^m$ (with ${\bf x}$ being Lorentz products of $e$ and $p$), and the mixed amplitudes are given by corresponding $m$-th order derivatives, which are known as transmuted operators~\cite{Cheung:2017ems} acting on the full amplitude, $\frac{\partial^m}{\partial {\bf x}^m} A({\bf x})$. 


{\bf Examples} Before proceeding, we present a few simple examples for these expansions. For $n=3$ YM amplitude, the expansion is trivial: we have two terms,  $e_1\cdot e_3 A^{\mathrm{YM} \oplus \phi^3}(\{2\}|1,3)=\frac{1}{2}e_1 \cdot e_3 e_2\cdot (p_1-p_3)$ with $r=0$, and $-e_1\cdot f_2 \cdot e_3$ with $r=1$ (where $A^{\phi^3} (1,2,3)=1$). For $n=4$ we have terms with $r=0,1,2$:
\begin{eqnarray}
    A^\mathrm{YM}_4&=&\frac{1}{2!}e_1\cdot e_4 A^\mathrm{YM\oplus \phi^3}(\{2,3\}|1,4)\\
    &-&e_1\cdot f_3\cdot e_4 A^\mathrm{YM\oplus \phi^3}(\{2\}|1,3,4)\nonumber\\
    &+&e_1\cdot f_2\cdot f_3\cdot e_4 A^\mathrm{\phi^3}(1,2,3,4)+(2\leftrightarrow 3)\nonumber
\end{eqnarray}
and exactly the same expansion holds with YM ($\phi^3$) replaced by GR (YM) or BI (NLSM). For $n=4$ EFT amplitudes, we have only one term with $r=1$: $A_4^\mathrm{NLSM} =p_2 \cdot \hat{f}_3 \cdot p_4 A_4^{\phi^3} (1234)=s_{12}+s_{23}$, and for $n=6$ we have
\begin{eqnarray}
A^\mathrm{NLSM}_6&=&
    p_2\cdot \hat{f}_3\cdot \hat{f}_4\cdot \hat{f}_5\cdot p_6 A^\mathrm{\phi^3}(1,2,3,4,5,6)
    \\
    \nonumber &+&\frac{1}{2!}
    p_2\cdot \hat{f}_3\cdot p_6 A^\mathrm{NLSM\oplus \phi^3}(\{4,5\}|1,2,3,6)\\
    &+&\mathrm{Perm}(3,4,5). \nonumber
\end{eqnarray}
Again the same holds when NLSM ($\phi^3$) is replaced by sGal (NLSM), BI (YM) and DBI (YMs). 
\section{Universal expansions for gauge theories and gravity}
In this section we derive expansions of GR and YM amplitudes~\footnote{Such an argument does not directly apply to expansions of BI amplitudes due to the lack of such uniqueness theorem for the latter.} from the powerful uniqueness theorem based on gauge invariance~\cite{Rodina:2016jyz,Arkani-Hamed:2016rak}. We start by writing general forms of the ansatz:
\begin{equation}
    A_n=\sum_g \frac{N_g}{\prod_{i} P_{g,i}},
\end{equation}
where we sum over all possible cubic graph $g$ (only planar ones for color-ordered YM amplitudes), and for each graph $g$ we have $n{-}3$ propagators ($i=1,\cdots, n{-}3$). The numerator takes the form
\begin{equation*}
    N_g=\sum_I m_I c_{g,I},
\end{equation*}
where we sum over a basis of monomials $m_I$ of Lorentz products $e \cdot e, e \cdot p, p \cdot p$ with constants $c_{g,I}$ to be fixed. Each monomial contains $s$ copies of polarization vectors $e_i$ for $i=1,2,\ldots,n$, and $k=s (n{-}2)$ powers of momenta from power counting ($s=1,2$ for YM and GR respectively). The basis is determined from constraints $p_{i}^{2} = 0, e_{i}^{2} = 0, p_{i} \cdot e_{i} =0, \sum_{i=1}^{n} p_{i} = 0$; 
in particular, it can be chosen by eliminating $p_i \cdot p_n$ for $i\neq n$ and $p_1 \cdot p_{n{-}1}$, as well as $p_n \cdot e_i$ (for $i\neq n$), $p_1 \cdot e_{n}$. The ansatz for $n=4$ YM case 
\begin{equation*}
A_4=\frac{c_{1,1} e_1\cdot e_4 e_2\cdot p_1 e_3\cdot p_1}{s_{1,2}}+\frac{c_{2,1} e_1\cdot e_4 e_2\cdot p_1 e_3\cdot p_1}{s_{2,3}}+\cdots
\end{equation*}
has two planar cubic trees and $30$ monomials in our basis, thus there are $60$ constant parameters $c_{1,i}$ and $c_{2,i}$ for $i=1,\cdots, 30$~\footnote{Some of these parameters turn out to be redundant when Mandelstam variables in numerators and denominators cancel; {\it e.g.} only $57$ of these $60$ $c$'s are independent.}. For GR we have $3$ cubic trees and each numerator has $30^2$ monomials as ``square" of the YM ones, thus $2700$ parameters in total. 

Let us first review the {\it uniqueness theorem} for YM and GR amplitudes based on gauge invariance~\cite{Rodina:2016jyz}. It states that gauge invariance for any $n{-}1$ particles uniquely fixes the above ansatz $A_n$ (up to an overall constant) to be the correct $n$-point YM or GR amplitudes. The key in the proof of this theorem relies on the following {\it Lemma} which we will use shortly. 

Lemma: Let $B(k)$ be a polynomial linear in each polarization vector, with at most $k$ factors of the form $e \cdot p$ in any given term, then $B(k)$ can only be gauge invariant in at most in $k$ particles for $k<n{-}2$ (with momentum conservation in $n$ particles). With these at hand, we now move to our main claim:

{\bf Claim}: The gauge invariance in $n{-}1$ legs, {\it e.g.} $1,2,\cdots, n{-}1$, uniquely fixes the above ansatz $A_n$ (up to an overall constant) in the form of (\ref{expand one}), which in turn fix all mixed amplitudes contained in the expansion. 

We will prove this in three steps, and we focus on YM case (the proof for GR is completely analogous).

{\bf Step 1}: We show that after imposing gauge invariance of $\{2,3,\ldots,n-1\}$, each monomial of the ansatz must contain a Lorentz product of the form:
	\begin{equation}
		w(1,\alpha,n | \text{signs}) \equiv e_{1} \cdot v_{\alpha_1}   \left( \prod_{i=2}^{r} \bar{v}_{\alpha_{i-1}} \cdot v_{\alpha_i}   \right) \bar{v}_{\alpha_r} \cdot e_{n},
	\end{equation}
where we introduce a new notation: given an ordered set $\alpha=(\alpha_1, \cdots, \alpha_r)$, there are $2^r$ terms from $W(1, \alpha, n)$ and we label them with $r$ signs; we use the vector $v_i$ to denote either $p_i$ or $e_i$ for particle $i$, with $\bar{v}_i$ the other one, {\it i.e.} $(v_i, \bar{v}_i)=(p_i, e_i)$ or $(e_i, p_i)$, and the first/second choice is denoted by a $+$ or $-$ sign. For example, $w(1,2,3,4|-,+)=(e_1\cdot e_2) (p_2\cdot p_3) (e_3\cdot e_4)$ while $w(1,2,3,4|+,-)=(e_1\cdot p_2) (e_2\cdot e_3) (p_3\cdot e_4)$. In other words, using gauge invariance in $2, \cdots, n{-}1$ we will show that the ansatz takes the form:
\begin{equation} \label{pf1}
	A_n= \sum_{\alpha} \sum_{\text{signs}} w(1,\alpha,n | \text{signs}) C(1,\alpha,n | \text{signs}),
\end{equation}
where for each prefactor we denote its coefficients in the ansatz as $C$ which looks like an ``amplitude" since it again has all cubic trees with numerators given by remaining Lorentz products of $e$'s and $p$'s. 

To prove this, we look at the vector dotted into $e_1$ for each term in the ansatz. If it is $e_n$ we are done, and we only need to consider $e_i$ and $p_i$ for $i\neq 1,n$ in our basis. If $e_1$ is dotted with $p_i$, this term must contain $e_i$ and the chain goes on, but if we have $e_1 \cdot e_i$, then gauge invariance of particle $i$ forces us to have another term, with $e_1 \cdot p_i$, thus we have $A_n=e_1\cdot p_i C^+_i+ e_1 \cdot e_i C^-_i + \cdots$. Under the replacement $e_i \rightarrow p_i$, we see that the two terms must cancel against each other, which means that $C_i^-$ must contain $p_i$ in it (such that there is a chance for it to cancel $C_i^+|_{e_i \to p_i}$ with the replacement). We have shown that in our basis, when $e_1$ is dotted with $v_i$, there must be $\bar{v}_i$ in the same term, and we can continue in asking what vector is dotted into $\bar{v}_i$. In this way, we see that for any term we have a prefactor as
\begin{equation*}
    e_{1} \cdot v_{\rho_1}   \left( \prod_{i=2}^{t} \bar{v}_{\rho_{i-1}} \cdot v_{\rho_i}   \right)  \bar{v}_{\rho_t}^\mu
\end{equation*}
where $|\rho|=t<n{-}2$. Now there are three possibilities with the vector dotted into $ \bar{v}_{\rho_t}$:
\begin{enumerate}
    \item $e_n$, for which we are done.
    \item $e_j$ or $p_j$ for $j \notin \{1,\rho,n\}$, for which we keep going and extends the chain. 
    \item $p_j$ for $j \in \{1,\rho\}$ which is ruled out by the Lemma: note the coefficient of such a factor is equivalent to a polynomial $B(k)$ with at most $k=n{-}t{-}3$ $e \cdot p$ factors, but we require it to be gauge invariant in $n{-}t{-}2$ particles, and that leads to a contradiction. 
\end{enumerate}
This concludes our proof for Eq.\eqref{pf1}. 

{\bf Step 2}: To proceed, it is crucial to note that the replacement $e_{\alpha_j} \rightarrow p_{\alpha_j}$ eliminates the difference between  $v_{\alpha_j}$ and $\bar{v}_{\alpha_j}$, thus gauge invariance in this particle allows us to relate two terms in Eq.\eqref{pf1} with only one sign difference for $\alpha_j$; both terms contain a factor of the form
$$
    e_{1} \cdot v_{\alpha_1} \cdots  \bar{v}_{\alpha_{j-1}} \cdot p_{\alpha_j}
    \times  p_{\alpha_j}\cdot v_{\alpha_{j+1}} \cdots \bar{v}_{\alpha_r} \cdot e_{n}$$
and clearly the two coefficients only differ by a sign:
\begin{equation}
    C(1,\alpha,n|\ldots,+,\ldots)=-C(1,\alpha,n|\ldots,-,\ldots), 
\end{equation}
thus $2^r$ coefficients with the same $\alpha$ at most differ by a sign ($q$ denotes the number of $-$ in the signs):
\begin{equation}
    C(1,\alpha,n|\text{signs})=(-1)^q C(1,\alpha,n|\text{all plus}).
\end{equation}
Therefore, for each $\alpha$, the $2^r$ prefactors $w$ exactly combine into $W(1, \alpha, n)=\sum_{signs} (-1)^q w(1,\alpha,n|\text{signs})$ and we have
 \begin{equation}  \label{pf2}
    A_n= \sum_{\alpha} W(1,\alpha,n) A^{\prime}(\{\bar{\alpha}\} | 1,\alpha,n ).
\end{equation}
where we have defined the coefficients
\begin{equation}
A^{\prime}(\{\bar{\alpha}\}|1,\alpha,n )\equiv C(1,\alpha,n|\text{all plus}).
\end{equation}
Furthermore, we see that $A^{\prime}(\{\bar{\alpha}\}|1,\alpha,n )$ must be gauge invariant for particles in $\{\bar{\alpha}\}$ since each term does not talk to each other under gauge transformations of $\{2, \cdots, n{-}1\}$. We conclude that by even just gauge invariance of $n{-}2$ particles puts strong constraints on $A_n$, such that it takes the expansion form Eq.\eqref{pf2} with gauge invariant coefficients $A'$, which need to be fixed. 

Finally, we note that since each $A'$ is the coefficient of $w(1,\alpha,n | \text{all plus})$ of $A_n$, it can be extracted using a differential operator acting on $A_n$:
$$     A^{\prime}( \{\bar{\alpha}\} |1,\alpha,n )= \partial_{e_{1} \cdot p_{\alpha_1} }  \left( \prod_{i=2}^{r} \partial_{e_{\alpha_{i-1}} \cdot p_{\alpha_i} }  \right) \partial_{e_{\alpha_r} \cdot e_{n}} A_n.
$$


{\bf Step 3}: Before proceeding, we need one of the main results of \cite{Cheung:2017ems}, where {\it transmuted operators} for $\oplus$ was introduced. Single-trace $\text{YM} \oplus \phi^3$ ( $\text{GR} \oplus \text{YM}$) amplitudes with an ordered subset $\beta$ can be obtained by acting such operators on the YM/GR amplitude, {\it e.g.}
\begin{equation}
T\left[ \beta \right]  A^{\mathrm{YM}}=A^{\mathrm{YM \oplus \phi^3}}( \{\bar{\beta}\}|\beta ),
\end{equation}
where gluons in $\{\beta\}$ are transmuted into bi-adjoint $\phi^3$ scalars by the operator (we denote $s=|\beta|$)
\begin{equation}
    T[\beta] =  \partial_{e_{\beta_1} \cdot e_{\beta_s}} \prod_{i=2}^{s-1} \partial_{e_{\beta_i} \cdot (p_{\beta_{i-1}} - p_{\beta_s}) },
\end{equation}
An example is $ T[1,2,3,4]= \partial_{e_{1} \cdot e_{4}} \partial_{e_{2} \cdot (p_{1} - p_{4}) } \partial_{e_{3} \cdot (p_{2} - p_{4}) }$. 
A crucial point is that in our basis where $p_n$ is eliminated by momentum conservation, all the derivatives w.r.t. $e_i \cdot p_n$ can be dropped, thus the operator becomes very similar to the one extracting $A'$ from $A_n$!

As a final step, we impose gauge invariance on $1$ or $n$, which fixes $A_n$ to be $A_n^{\rm YM}$ thus all remaining parameters in those $A'$'s are completely fixed. Not surprisingly they turn out to be those mixed amplitudes we want:
\begin{equation}
\begin{split}
   A^{\prime}(\{\bar{\alpha}\}|1,\alpha,n ) &= T[\alpha^{-1},1,n] A^{\text{YM}}_n(1,2,\ldots,n)  \\
     & = A^{\text{YM}\oplus\phi^3}( \{\bar{\alpha}\} |\alpha^{-1},1,n )  \\
     & =(-1)^{r}   A^{\text{YM}\oplus\phi^3}( \{\bar{\alpha}\}|1,\alpha,n ),
\end{split}
\end{equation}
where we have used the fact that the amplitude picks up $(-1)^r$ under reflection, and we arrive at Eq.\eqref{expand one}.

We have also explicitly checked our proof for $n=4,5$. Starting from our ansatz with $60$ parameters, by imposing gauge invariance of particle $2,3$, we indeed find the expansion form (\ref{pf2}) and only $9$ parameters remain for those $A'$. 
By imposing gauge invariance of $1$ or $4$, the amplitude becomes Eq.(\ref{expand one}) up to an overall constant. For $n=5$, the general ansatz includes 2475 parameters, imposing gauge invariance of $2,3,4$ gives Eq.(\ref{pf2}) with only 72 parameters left; gauge invariance for $1$ or $5$ uniquely fixes it as the expansion of $n=5$ YM amplitudes. 

%
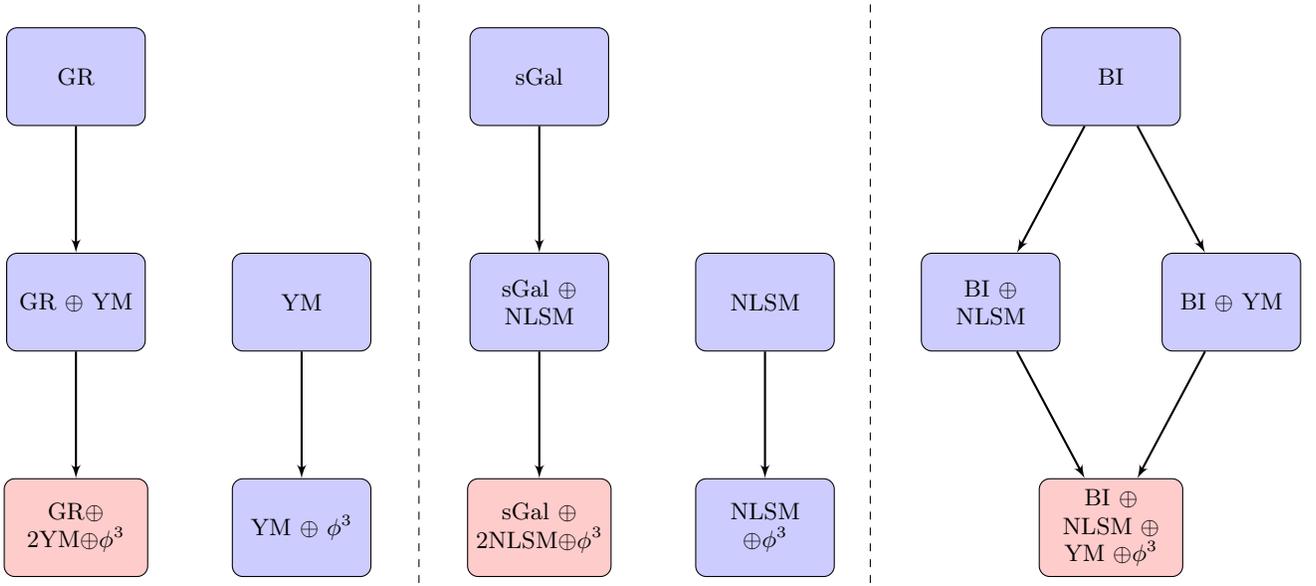
\begin{figure*}
\begin{center}

\tikzstyle{block} = [rectangle, draw, fill=blue!20, text width=5em, text centered, rounded corners, minimum height=4em]
\tikzstyle{block1} = [rectangle, draw, fill=red!20, text width=5.2em, text centered, rounded corners, minimum height=4em]
\tikzstyle{line} = [draw, thick, -latex']
\tikzstyle{cloud} = [draw, ellipse, fill=red!20, node distance=2.7cm, minimum height=2em]

\begin{tikzpicture}[node distance = 3cm, auto, scale=.8]
\begin{scope}[xshift=-.7cm]
    \node [block] (gr) {GR};
    \node[block, below of= gr] (eym) {GR $\oplus$ YM};
    \node [block1, below of= eym] (eyms) {GR$\oplus$ $2$YM$ \oplus\phi^3$};
    \path [line] (eym) --node  {} (eyms);
    \path [line] (gr) --node  {} (eym);
    \node [block, right of=eym] (ym) {YM};
    \node[block, below of= ym] (yms) {YM $\oplus~\phi^3$};
    \path [line] (ym) --node [pos=.5] {} (yms);
\end{scope}
%
%
\draw[dashed] (5,1.2)--(5,-8.5);
\begin{scope} [xshift=7cm]
    \node [block] (sgal) {sGal};
    \node[block, below of= sgal] (sgalnlsm) {sGal $\oplus$ NLSM};
    \node [block1, below of= sgalnlsm] (nlsms) {sGal $\oplus$ $2$NLSM$\oplus\phi^3$};
    \path [line] (sgalnlsm) --node {} (nlsms);
    \node[block, right of= sgalnlsm] (nlsm) {NLSM};
    \node[block, below of=nlsm] (phi3) {NLSM $\oplus \phi^3$};
    \path [line] (sgal) --node {} (sgalnlsm);
    \path [line] (nlsm) --node {} (phi3);
\end{scope}
\draw[dashed] (12.5,1.2)--(12.5,-8.5);
\begin{scope}[xshift=16.5cm]
    \node [block] (bi) {BI};
    \node[block, below of=bi,xshift=-1.6cm] (ym1) {BI $\oplus$ NLSM};
    \node [block1, below of= ym1,xshift=1.6cm] (dbis) {BI $\oplus$ NLSM $\oplus$ YM $\oplus \phi^3$};
    \path [line] (ym1) --node {} (dbis);
    \node[block, below of=bi,xshift=1.6cm] (ym2) {BI $\oplus$ YM};
    \path [line] (bi) --node   {} (ym1);
    \path [line] (bi) --node {} (ym2);
    \path [line] (ym2) --node {} (dbis);
\end{scope}
\end{tikzpicture}
\caption{Summary of universal expansions}
\label{fig:summary}
\end{center}
\end{figure*}
\section{Universal expansions for effective field theories}
The expansions of EFTs can be obtained by imposing the special DR~\cite{Cheung:2017ems,Cheung:2017pion} on both sides of Eq.(\ref{expand one}). We define $(2d{+}1)$-dimensional momenta for the original particles to be $p_j=(p^\mu_j,0,0^\mu)$ with the first and third entries $d$-dimensional and the middle entry one-dimensional. We choose two special legs $a,b$ with polarization vector $e_a=e_b=(0^\mu, 1, 0^\mu)$; the remaining $n{-}2$ polarizations are chosen as $e_j=(p^\mu_j,0,i p^\mu_j)$ for $j\neq a, b$. Reducing to $d$ dimensions gives ($p_i \cdot p_j$ trivially reduce):
\begin{equation} \label{DR}
    e_{i} \cdot e_{j} \rightarrow \begin{cases} 1 & \{i, j\}=\{a,b\} \\
    0 &  \text { otherwise }\end{cases}, \
    e_{i} \cdot p_{j} \rightarrow \begin{cases}
    0 & i \in\{a,b\} \\
    p_{i} \cdot p_{j} & \text { otherwise }
    \end{cases}
\end{equation}
It is remarkable that gluons in $2d{+}1$ dimensions then become pions in $d$ dimensions; equivalently, we can use the operators that transmute gluons to pions~\cite{Cheung:2017ems}:
\begin{equation}
        \frac{\partial}{\partial\left(e_{a} e_{b}\right)} \prod_{i\neq a,b}^{n}\left(\sum_{j \neq i}^n p_{i} p_{j} \frac{\partial}{\partial\left(p_{j} e_{i}\right)}\right)  A^\text{YM}_n=
        A^\text{NLSM}_n.
\end{equation}
By applying DR or such operators on RHS of Eq.(\ref{expand one}), they transmute {\it e.g.} gluons in $\{\bar{\alpha}\}$ into pions term-by-term and change the coefficients $W$ to $\hat{W}$. It is interesting that different choices of $a,b$ (w.r.t. $1,n$) lead to three different types of EFT expansions, and we keep in mind that mixed amplitudes with odd number of Goldstone particles in theory $I$ vanish. 

If we choose $a,b=1,n$, then on the RHS only the term with $e_1 \cdot e_n\to 1$ survives from DR and we reach at the well-known fact that the amplitude in theory $I$ is identical to that with only two particles in $II$ (as a trivial expansion) ${A}_n^{\uppercase\expandafter{\romannumeral1}}=A^{\uppercase\expandafter{\romannumeral1} \oplus \uppercase\expandafter{\romannumeral2}}(\{2,\cdots,n-1\}|1,n )$~\cite{Cachazo:2016njl}. 

The second choice is {\it e.g.} $a,b=1,2$, then we reach at the expansion in Eq.(\ref{expand two}), where we must have odd $r$ since we need the number of particles in $I$, $r{+}3$, to be even, and we have $\sum _{i=1,\text{odd}}^{n-3} i! \binom{n-3}{i}$ terms on the RHS. Note the only non-vanishing term must start with $1, 2$ and since $e_1 \cdot e_2 \to 1$ we have $p_2$ on the left end of $\hat{W}$.

Finally, we have the slightly more complicated expansion with the third choice {\it e.g.} $a,b=2,3$:
\begin{eqnarray}    
    {A}_n^{\uppercase\expandafter{\romannumeral1}}&=&\sum_{\alpha,\beta} (-1)^{s+1} (p_1\cdot \hat{f}_{\alpha_1}\cdots \hat{f}_{\alpha_r}\cdot p_{2}) (p_{3} \cdot \hat{f}_{\beta_1}\cdots \hat{f}_{\beta_s} \cdot p_{n})  \nonumber
    	\\ 
    	&\times& A^{\uppercase\expandafter{\romannumeral1} \oplus \uppercase\expandafter{\romannumeral2}}( \{\bar{\alpha}\cap\bar{\beta}\}|1,\alpha,2,3,\beta,n )+(2 \leftrightarrow 3),
    	\label{expand two2}
\end{eqnarray}
where we sum over (non-intersecting) ordered sets $\alpha, \beta \subset \{4, \cdots, n{-}1\}$ with  $r=|\alpha|,s=|\beta|$ and $r{+}s$ even; particles in $II$ form an ordering $(1, \alpha, 2, 3, \beta, n)$, and we refer to the complementary set with $n{-}4{-}r{-}s$ particles in $I$ as $\{\bar{\alpha}\cap\bar{\beta}\}$. One can check that we have $\sum _{i=0, \text{even}}^{n-4} 2 (i+1) i!  \binom{n-4}{i}$ terms in the expansion. Even for $n=4$, we have two terms 
$A^\text{NLSM}_4=-p_1 \cdot p_2 p_3\cdot p_4 A^{\phi^3}(1,2,3,4)+(2\leftrightarrow 3)$. These expansions hold for even $n$, and when $n$ is odd, the EFT amplitudes vanish and instead we have a non-trivial linear relation for mixed amplitudes from DR. Such relations take the same form as Eq.(\ref{expand two}) and Eq.(\ref{expand two2}) with the only difference that $r$ or $r+s$ are even or odd, in these two cases. 

Last but not least, there are different ways of applying DR to the GR expansion, which lead to expansions for other EFTs. Note that the polarization tensor contains two sets of polarization vectors $e, \tilde{e}$, and in Eq.(\ref{expand one}) we have only expanded with $e$ in $W(1,\alpha,n)$ ($\tilde{e}$ side is untouched). If we perform DR with the replacement Eq.(\ref{DR}) on $e$, we have EFT expansion of BI (into BI$\oplus$YM); if we perform DR on $\tilde{e}$, we have gauge-theory expansion of BI (into BI$\oplus$NLSM); very nicely, if we perform DR on both $e$ and $\tilde{e}$, we obtain the (EFT) expansion of sGal into sGal $\oplus$ NLSM. All these expansions are summarized in Fig.\ref{fig:summary}.

\section{Double \& recursive expansions, and double copy}
In this section, we present more expansions, such as ``double expansions" in the case of GR, sGal and especially BI; we point out that our expansions represent the first step of expanding amplitudes in a recursive way, which allows us to connect to the double copy directly. 

Let us start with GR case, where we apply \eqref{expand one} to both $e$ and $\tilde{e}$, and this leads to a double expansion
\begin{eqnarray} \label{double expand GR}
A^\mathrm{GR} &=& \sum_{\alpha,\beta} (-1)^{r+s}  W(1,\alpha,n) \tilde{W}(1,\beta,n) \\
& \times& A^{\mathrm{GR  \oplus 2YM \oplus \phi^3}}(\{ \bar{\alpha} \cap \bar{\beta} \}| \alpha \cap \bar{\beta} | \beta \cap   \bar{\alpha} |1,\alpha \cap \beta,n  )\nonumber
\end{eqnarray}
where the prefactor $W(1,\alpha,n)$ and $\tilde{W}(1,\beta,n)$ contains $e$ and $\tilde{e}$ respectively; the four subsets of particles refer to $\phi^3$ scalars (with both orderings), gluons with ordering $\alpha$ (polarization $\tilde{e}$) and $\beta$ ($e$), as well as gravitons. By applying DR on both $e$ and $\tilde{e}$ gives double expansion for sGal, and if we only apply it on $e$ or $\tilde{e}$, we have the double expansion for BI (with poins, gluons and $\phi^3$ scalars in mixed amplitudes). All these are consequences of \eqref{expand one} and \eqref{double expand GR} which originate from the uniqueness theorem; alternatively they can be derived from double copies of ``basic" expansions of YM and NLSM amplitudes, which implies that $\oplus$ and $\otimes$ ``commute", {\it e.g.} $({\rm YM}\oplus\phi^3)\otimes ({\rm NLSM} \oplus \phi^3)={\rm BI}\oplus {\rm NLSM} \oplus {\rm YM} \oplus \phi^3$. We summarize them with red color in Fig.\ref{fig:summary}.

What is more interesting is that these mixed amplitudes can be expanded further with more particles in $II$, leading to a recursive expansion of the original amplitude in $I$. At each step we need to pick a ``reference particle", {\it e.g.} for EFTs, \eqref{expand two} can already be viewed as expanding the (trivial) mixed amplitude $A^{I\oplus II}_n(\{2,\cdots, n{-}1\}|1,n)$ with reference $2$. We focus on gauge-theory case since EFT ones can be derived via DR, whose precise form depends on the choice of $a,b$. Quite nicely, such recursive expansions for gauge theories and gravity read~\cite{Fu:2017uzt,Du:2017kpo,Teng:2017tbo,Du:2017gnh}:
\begin{equation}
\begin{split}
	&A^{\uppercase\expandafter{\romannumeral1} \oplus \uppercase\expandafter{\romannumeral2}}( \{\bar{\alpha}\}|1,\alpha,n )= \sum_{\beta,\shuffle,j}   e_{\bar{\alpha}_0} \cdot f_{\beta_1} \cdot f_{\beta_2} \ldots  f_{\beta_s} \cdot p_{\alpha_j} \times  \\
    & A^{\uppercase\expandafter{\romannumeral1} \oplus \uppercase\expandafter{\romannumeral2}}(\{\bar{\beta}\}|1,
    \cdots,\alpha_j,(\alpha_{j+1},\cdots,\alpha_r)\shuffle (\beta^{-1}, \bar{\alpha}_0),n ),
\end{split}
\end{equation}
where $\bar{\alpha}_0$ is an arbitrary reference particle in $\{\bar{\alpha}\}$, $\beta$ is an ordered subset with $s$ labels of $\{\bar{\alpha}\}/\{\bar{\alpha}_0\}$ and $\{\bar{\beta}\}$ is the complementary set as usual; we need to sum over $j=0,1,\ldots,r$ (with $\alpha_0 \equiv 1$ for $j=0$ case), and shuffle the two ordered sets. For example, with reference $2$, $A^{YM\oplus \phi^3}(\{2,3,4\}|1,5,6)$ can be expanded as \begin{equation}   \begin{split}       &\frac{1}{2!}e_2\cdot p_1 A^{YM\oplus \phi^3}(\{3,4\}|1,(5)\shuffle (2),6) \\       &+\frac{1}{2!}e_2\cdot p_5 A^{YM\oplus \phi^3}(\{3,4\}|1,5,2,6) \\        &+e_2\cdot f_3  \cdot p_1 A^{YM\oplus \phi^3}(\{4\}|1,(5)\shuffle (3,2),6) \\&+e_2\cdot f_3\cdot p_5 A^{YM\oplus \phi^3}(\{4\}|1,5,3,2,6) \\   &+e_2\cdot f_3\cdot f_4\cdot p_1 A^{ \phi^3}(1,(5)\shuffle (4,3,2),6) \\   &+e_2\cdot f_3\cdot f_4\cdot p_5 A^{ \phi^3}(1,5,4,3,2,6) \\   &+\mathrm{Perm(3,4)}    \end{split} \end{equation}  
One can keep going until only $A^{\phi^3}$ amplitudes remain on the RHS. In general, the end result for such recursive expansion reads
$
A_n^I=\sum_{\pi \in S_{n{-}2}} N_n^{I/II} (\pi)~ A_n^{II}(1, \pi, n)$
where we have expressed $A^I_n$ as a linear combination of $(n{-}2)!$ ordered amplitudes in II with coefficients $N_n$ known as the {\it BCJ master numerators} in the theory $I/II$ (such that $I/II \otimes II=I$). Note that the ``quotient" theory is again universal: for gauge-theory expansions $I/II={\rm YM}$, and for EFT ones, $I/II={\rm NLSM}$. Such expansions thus provide a systematic way for extracting kinematic numerators needed in all these theories, and this way of extracting them was originally found using CHY formulas in~\cite{Du:2016tbc,Fu:2017uzt, Du:2017kpo} (see also \cite{Mafra:2011kj,Carrasco:2016ldy}) and automatized in~\cite{Edison:2020ehu,He:2021lro}. Different choices of reference particles lead to different recursive expansions and BCJ numerators, but all of them are equivalent. It is an interesting open question if we can derive such expansions directly from {\it e.g.} gauge invariance and Adler zero.

\section{Conclusion and Discussions}
In this note we study certain expansions of amplitudes which work universally in gauge theories, gravity and various EFTs, and they ``interpolate" $\oplus$ and $\otimes$ operations that connect all these amplitudes. While gauge-theory expansions follow from uniqueness based on gauge invariance, currently we have only derived expansions in EFTs (including BI) from the gauge-theory ones via DR; their Adler-zero uniqueness relies on singularity structure of quartic diagrams but individual mixed amplitudes involve cubic diagrams, thus more works are needed for obtaining them directly from uniqueness. In addition, we do not know how to derive recursive expansions, which eventually lead to BCJ numerators of YM/NLSM, without referring to CHY formulas. We would like to understand all these from the perspective of constraining amplitudes from basic principles. 

Relatedly, one can prove such expansions by using on-shell recursion relations~\cite{Britto:2004ap, Britto:2005fq, Cheung:2015ota}, since it is straightforward to show that residues at any poles agree on both sides by factorizations. What remains to show is the absence of pole at infinity, {\it e.g.} for BCFW shifts of $1,n$ in gauge-theory cases. Note that gauge invariance for $n{-}2$ particles are manifest in \eqref{expand one}, and the non-trivial point is the gauge invariance in $1$ or $n$, which in turn implies such behavior at infinity~\cite{Rodina:2016mbk}. A similar argument applies to EFT expansions \eqref{expand two} where we need to show (enhanced) Adler zero of $1,2,n$, which also relates to the behavior at infinity~\cite{Carrasco:2019qwr}. It would be highly desirable to understand all these better. Another interesting question concerns possible relations of our EFT expansion with the appearance of these mixed amplitudes from soft limits, where they were first discovered as ``extensions" of original theories~\cite{Cachazo:2016njl}. Of course all relations we discussed follow from CHY formulas, and as usual we can look for their origins in string theory ({\it c.f.}~\cite{Carrasco:2016ldy, Carrasco:2016ygv, Stieberger:2009hq, Bjerrum-Bohr:2009ulz, Bjerrum-Bohr:2010pnr, Stieberger:2016lng, Schlotterer:2016cxa}). The question we 
ask here is, however, can we demystify these relations purely from field-theory perspective (see~\cite{Cheung:2016prv,Cheung:2021zvb})?

Last but not least, a direct consequence of Eq.~(\ref{expand one}) is a similar expansion for corresponding one-loop integrands obtained by the forward limit~\cite{Caron-Huot:2010fvq,He:2015yua,Cachazo:2015aol}, which is in the representation naturally given by ambitwistor string~\cite{Mason:2013sva, Adamo:2013tsa} at one loop~\cite{Geyer:2015bja, Geyer:2015jch} (see also~\cite{He:2017spx,Edison:2020uzf}). As an example, we first expand the $(n+2)$-pt  YM tree amplitude:
\begin{equation}
\begin{split}
A^\mathrm{YM}&(+,1,2,\ldots,n,-)= \\
   &\sum_{\alpha}(-1)^r W(+,\alpha,-) A^\mathrm{YM \oplus \phi^3} (\{\bar{\alpha}\} |+,\alpha,- ), 
   \end{split}
\end{equation}
where we expand the amplitude by fixing external legs labeled by $+, -$, and $\alpha$ denotes ordered subsets of \{1,2,\ldots,n\}. We take the forward limit of $+,-$ on both sides, identify $e_+$ with $e_-$ and sum over the possible states; the prefactor becomes

\begin{equation*}
   W(+,\alpha,-) \ {\raisebox{.5em}{$\underrightarrow{\text{ F.L. }}$}} \ \operatorname{Tr}( f_{\alpha_1} \cdot f_{\alpha_2} \ldots  f_{\alpha_r}).
\end{equation*}

Thus we get the one-loop expansion for YM:
\begin{eqnarray}
&&A^{\text{YM,loop}}(1,2,\ldots,n) =\\
&& \sum_{\alpha} (-1)^r \operatorname{Tr}( f_{\alpha_1} \cdot f_{\alpha_2} \ldots  f_{\alpha_r})
     A^\mathrm{YM \oplus \phi^3,scalar-loop}( \{\bar{\alpha}\}|\alpha).\nonumber
\end{eqnarray}
It would be interesting to further study such expansions in gauge theories and EFTs at loop level. 

\begin{acknowledgments}
We would like to thank Bo Feng and Yong Zhang for helpful discussions. The research of S. H. is supported in part by National Natural Science Foundation of China under Grant No.11935013, 11947301, 12047502, 12047503.
\end{acknowledgments}

\bibliography{apssamp}

\end{document}